%% file: arxiv.tex
\documentclass[preprint,review,12pt]{elsarticle}

\usepackage{graphicx}
\usepackage{braket}
\usepackage{amssymb}
\usepackage{amsmath}

\usepackage{dcolumn}
\usepackage{bm}
\usepackage{siunitx}
\usepackage[colorlinks=false, pdfborder={0 0 0}]{hyperref}
\usepackage{booktabs}
\usepackage{braket}
\usepackage{threeparttable}
\usepackage{tabularx}
\usepackage{float}
\usepackage{amsmath}

\usepackage{lineno}

\biboptions{sort&compress}

\journal{Physics Letters B}

\begin{document}

\begin{frontmatter}


\title{Reconciling anomalous quadrupole deformation in $^{200,202}$Hg with Coulomb excitation}



\author[Surrey]{G.~Willmott}\ead{gw00431@surrey.ac.uk}

\author[Surrey,NPL]{J.~Heery}

\author[Surrey]{J.~Henderson}\ead{jack.henderson@surrey.ac.uk}

\author[Surrey]{P.~Stevenson}

\author[Surrey]{C. Cousins}

\author[Surrey]{D.~T.~Doherty}

\author[ANL]{A.~Ertoprak}

\author[ANL]{B.~P.~Kay}

\author[Surrey]{L.~Morrison}

\author[ANL]{C.~M\"uller-Gatermann}

\author[Surrey]{C.~O'Shea}

\author[Surrey,NPL]{P.~H.~Regan}

\author[ANL]{W.~Reviol}

\author[TRIUMF,LLNL]{D.~M.~Rhodes}

\author[Surrey]{R.~Russell}

\author[ANL]{N.~Sensharma}

\author[ANL]{D.~Seweryniak}

\author[ANL]{M.~Siciliano}

\author[Surrey]{R.~S.~Sidhu}

\author[ANL]{I.~A.~Tolstukhin}

\author[LLNL]{C.~Y.~Wu}

\address[Surrey]{School of Mathematics and Physics, University of Surrey, Guildford GU2 7XH UK}
\address[NPL]{National Physical Laboratory, Hampton Road, Teddington, Middlesex, TW11 0LW, UK}
\address[ANL]{Physics Division, Argonne National Laboratory, Lemont, Illinois 60439, USA}
\address[LLNL]{Lawrence Livermore National Laboratory, Livermore, California 94550, USA}
\address[TRIUMF]{TRIUMF, Vancouver, British Columbia V6T 2A3, Canada}


\input{abstract}


\end{frontmatter}

\input{content}

\bibliographystyle{elsarticle-num} 

\bibliography{bibliography}

\end{document}

%% file: abstract.tex
\begin{abstract}
Stable isotopes of mercury, with two fewer protons than magic lead, are found experimentally to exhibit oblate deformation, similar to neighbouring platinum isotopes. These isotopes are of special interest due to the use of $^{199}$Hg as an experimental laboratory for electric dipole moment searches, which are a powerful test of charge-parity ($CP$) violation and the interpretation of which depends critically on the quadrupole and octopule shapes of the underlying nucleus. It is observed that the ratio between experimental quadrupole moments and electric quadrupole transition strengths exceed physical limits, placing uncertainty on this underlying shape. In an effort to resolve this situation, a Coulomb excitation experiment was performed to remeasure the spectroscopic quadrupole moment ($Q_s(2^+_1)$) of $^{200}$Hg and $^{202}$Hg utilising $^{48}$Ti and $^{120}$Sn targets. The values obtained were $Q_s(2^+_1)$ $^{200,202}$Hg = $0.52(14)$, $0.63(10)$ eb, lying within physical limits defined by $B(E2)$ values but still consistent with near axially-symmetric oblate deformation. A systematic study of nuclei indicates that, as with neighbouring Pt isotopes, $^{200,202}$Hg are strongly oblate deformed, albeit without fully developing into an axial rotor.
\end{abstract}

%% file: content.tex


\def\approxprop{%
  \def\p{%
    \setbox0=\vbox{\hbox{$\propto$}}%
    \ht0=0.6ex \box0 }%
  \def\s{%
    \vbox{\hbox{$\sim$}}%
  }%
  \mathrel{\raisebox{0.7ex}{%
      \mbox{$\underset{\s}{\p}$}%
    }}%
}

\section{Introduction} 

The magnitude and form of deformation of a nucleus emerges from its underlying nuclear structure. Quadrupole deformation is ubiquitous across the nuclear landscape~\cite{ref:Pritychenko_16} with significant variations in both magnitude and resultant shape. Mapping the emergence of different forms (i.e., prolate or oblate) of deformation thus informs our understanding of the microscopic properties of the nucleus. Of particular interest to this work, the spectroscopic quadrupole moment $Q_s$ is a measure of how nuclear charge deviates from sphericity. A positive $Q_s$ indicates oblate quadrupole deformation. By comparison, reduced electric-quadrupole transition strengths, $B(E2)$ values, relate to the magnitude of this deviation.

The $A\approx200$ region of the nuclear chart has limited $Q_s(2^+_1)$ data, and warrants special interest due to the proximity to doubly magic $^{208}$Pb and a region of shape coexistence in the neutron deficient Hg isotopes \cite{ref:Bockisch_79,ref:Olaizola_19}. These nuclei are also a rare example of a region of the nuclear landscape that exhibits extensive oblate quadrupole deformation. In addition, this region of nuclear landscape plays a critical role as a laboratory for investigations of fundamental symmetries, through electric dipole moment (EDM) searches in \textsuperscript{199}Hg.

A permanent EDM is a constant time-averaged separation of electrical charges in a system \cite{ref:Chupp_19}. The search for a non-zero permanent electric dipole moment remains a promising route to investigate beyond Standard Model physics \cite{ref:Butler_19,ref:Griffith_09}. Whilst combined Charge conjugation Parity Time reversal ($CPT$) symmetry is held as a good symmetry, $CP$ and thereby $T$ symmetry violation has previously been observed \cite{ref:Christenson_64,ref:Schubert_14}. Permanent EDMs challenge $T$ symmetry as a $T$ reversal transformation induces a spin reversal but the charge imbalance is by definition permanent and so cannot flip to produce a good time reversed partner to the original system \cite{ref:Ramsey_82}. Without $T$ violating extensions to the standard model the magnitude of the EDM must tend to zero, hence any measurement of a non-zero EDM would be a clear indication of new physics. These symmetry violating extensions may hold the key to understanding the departure from matter-antimatter symmetry in the universe \cite{ref:Gavela_94, ref:Gogberashvili_25, ref:Sakharov_67}. Presently, $^{199}$Hg is the system with the smallest measured limit of the size of an EDM \cite{ref:Griffith_09}. Previous EDM magnitude limits have been found for the electron $(<1.1\times10^{-29}$ e$.$cm$)$ \cite{ref:ACMECollaboration_18}, neutron $(<1.3\times10^{-26}$ e$.$cm$)$ \cite{ref:Abel_20}, deuteron $(<2.5\times10^{-17}$ e$.$cm$)$ \cite{ref:Andres_26}, and $^{199}$Hg atom $(<7.4\times10^{-30}$ e$.$cm$)$ \cite{ref:Graner_16}.

Permanent EDM studies of atoms rely on measuring the Schiff moment \cite{ref:Schiff_63}. The nuclear Schiff moment is approximately proportional to both the quadrupole and octupole deformation of the system~\cite{ref:Flambaum_20}:

\begin{equation}
    S \propto \beta_2 \beta_3^2.
    \label{eq:Schiff}
\end{equation}

Here $\beta_2$, $\beta_3$ are the dimensionless quadrupole and octupole deformation parameters respectively, and $S$ is the magnitude in the laboratory frame of reference. Thus, understanding both the octupole and quadrupole deformations within $^{199}$Hg will critically impact any conclusions from an EDM search. Note that $S$ has a linear dependence on $\beta_2$ and $\beta_3$ when in the intrinsic (rotating) reference frame \cite{ref:Flambaum_19}.

In spite of their importance to both fundamental symmetries investigations and their interest from a nuclear structure perspective, the stable Hg isotopes display some peculiarities that warrant further investigation. The observed $Q_s(2^+_1)$ values for both \textsuperscript{200,202}Hg exceed the limits expected of an axial rotor based on observed $B(E2)$ values. This situation cannot easily be rectified without the introduction of physical effects not anticipated in this mass region, such as significant clustering~\cite{ref:Mehl_25}. A discrepancy of 20\% arises between $\beta_2$ values extracted from the $B(E2)$ compared to with the $Q_s(2^+_1)$. Coulomb-excitation measurements utilising the reorientation effect are a common method to extract $Q_s(2^+_1)$~\cite{ref:Zielinska_22} and was the method used previously for $^{200}$Hg and $^{202}$Hg. The previous data have two issues. In addition to the apparently unphysical magnitude of $Q_s(2^+_1)$ in both nuclei, the two measurements in  \textsuperscript{202}Hg are discrepant by approximately 5$\sigma$ \cite{ref:Bockisch_79,ref:Spear_80}.

The aim of this work is to remeasure the $Q_s(2^+)$ of both $^{200}$Hg and $^{202}$Hg to resolve the aforementioned discrepancies with $Q_s(2^+_1)$. Doing so will allow the oblate shape of nuclei in this $A\approx200$ region to be confirmed, including $^{199}$Hg which will aid EDM measurements utilising this isotope.

\section{Experiment and Analysis}

The experiment was performed at Argonne National Laboratory at the Argonne Tandem Linear Accelerator System (ATLAS). Beams were provided to the target station at the end of the accelerator in Area 4. Samples of $^{200}$Hg and $^{202}$Hg were accelerated to 783.7 MeV (3.92 MeV/u and 3.88 MeV/u respectively) in a $31^+$ charge state. 

\begin{figure}[h]
    \centering
    \includegraphics[width=0.95\linewidth, trim=0.12cm 0.12cm 0.12cm 0.12cm, clip]{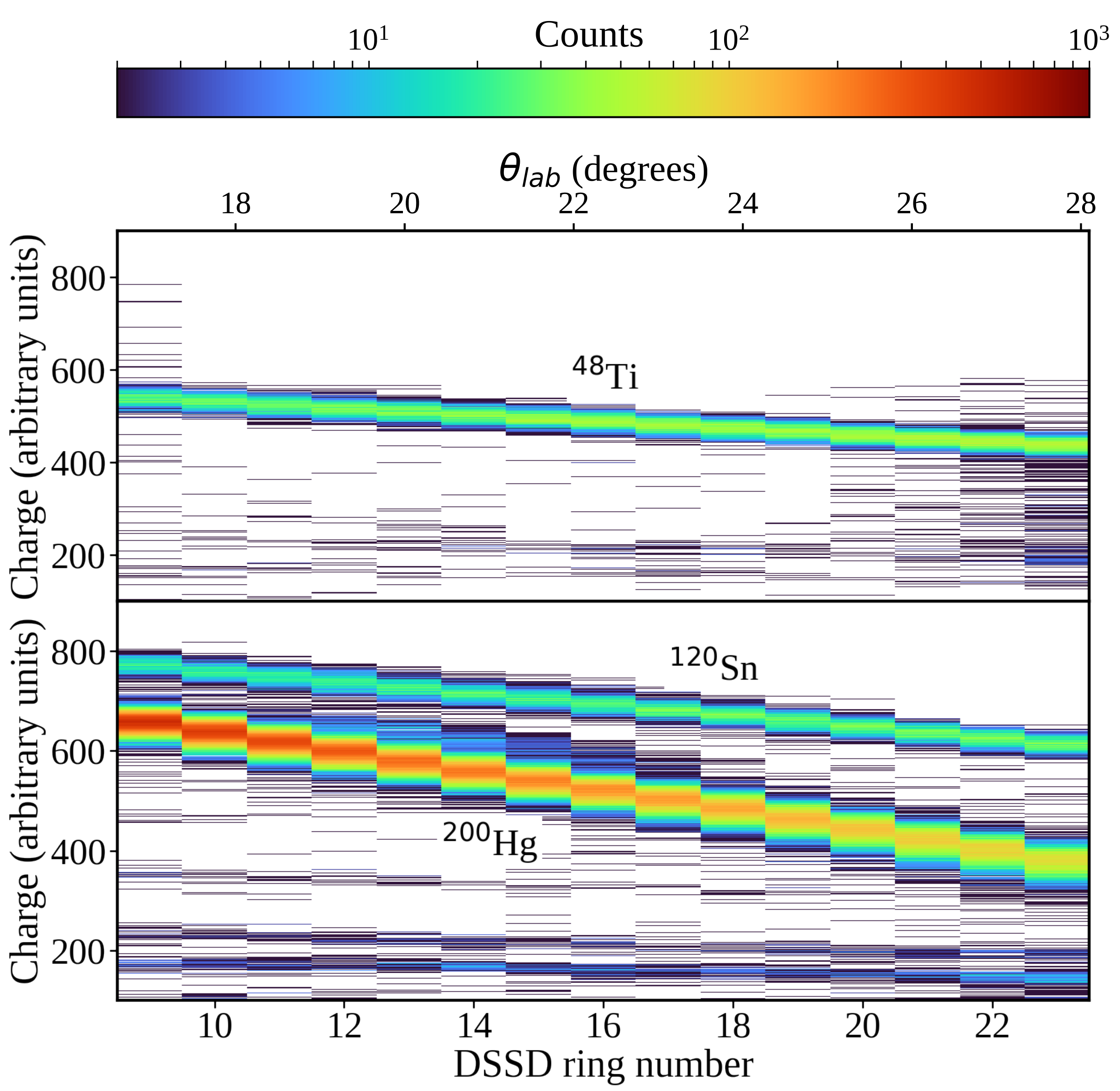}
    \caption{Particle kinematics as observed in the annular Si detectors: $^{200}$Hg impinged upon (top) $^{48}$Ti, and upon (bottom) $^{120}$Sn.}
    \label{fig:pid}
\end{figure}

\begin{figure}[h!]
    \centering
    \includegraphics[width=0.65\linewidth, trim=0.28cm 0.5cm 0.5cm 0.1cm, clip]{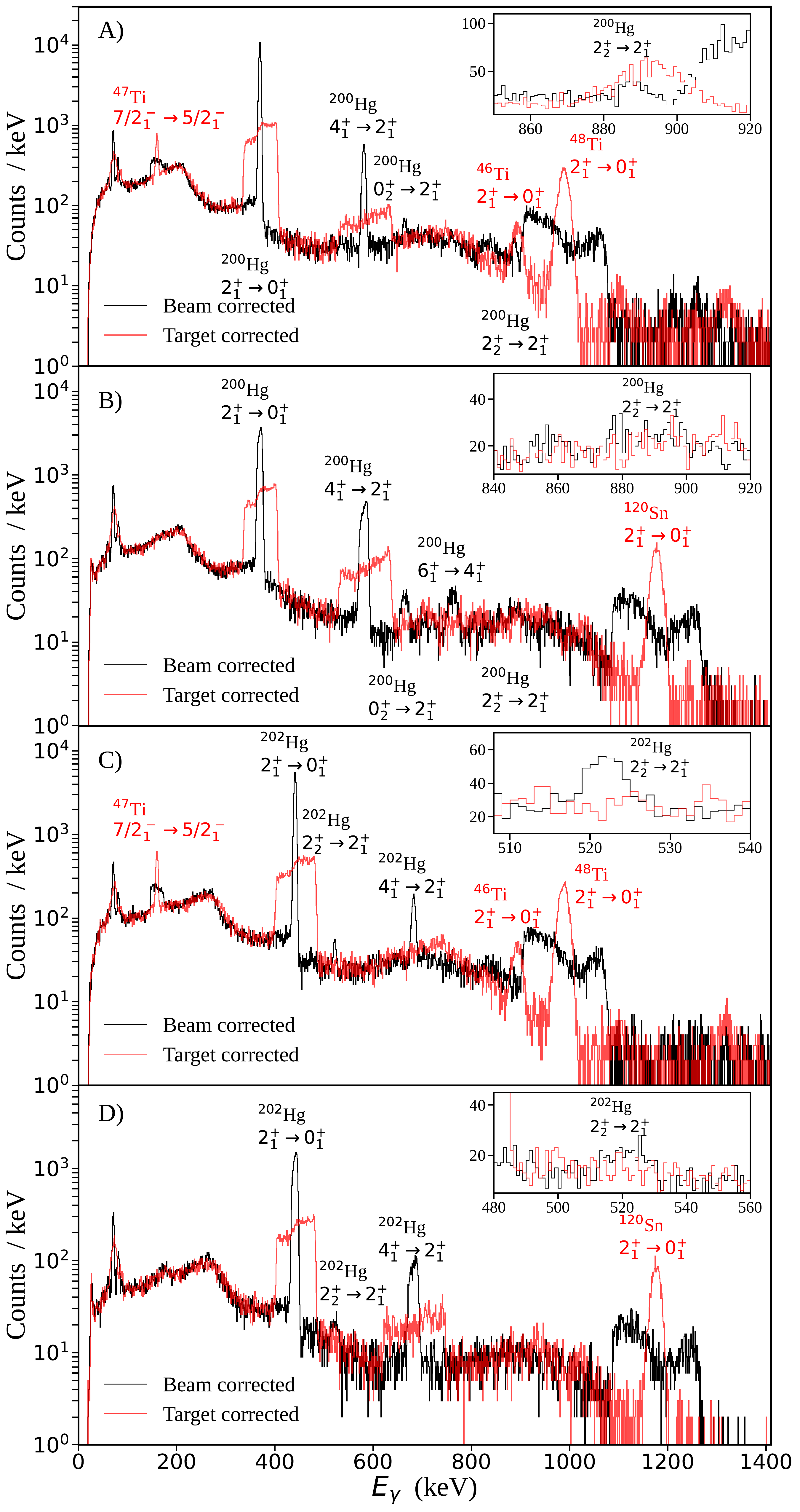}
    \caption{Doppler-corrected $\gamma$-ray spectra summed over all Gammasphere and all used DSSD angular range. (A) $^{200}$Hg impinged onto the $^{nat}$Ti target, (B) $^{200}$Hg onto $^{120}$Sn, (C) $^{202}$Hg onto $^{nat}$Ti, and (D) $^{202}$Hg onto $^{120}$Sn.}
    \label{fig:gammaspec}
\end{figure}

This is in line with the Cline criterion for ``safe'' Coulomb-excitation for the employed targets, in which the two nuclear surfaces maintain separation of at least 5 fm \cite{ref:Cline_86}. Beams with currents of approximately $6\times10^7$ pps (310 epA) were impinged on two targets; 452 $\mu$g$\cdot$cm$^{-2}$ of natural titanium (74\% $^{48}$Ti), and 780 $\mu$g$\cdot$cm$^{-2}$ of tin which was enriched to $>$98$\%$ $^{120}$Sn. Each beam-target combination was used for between 11 and 19 hours. Two targets were used to permit normalisation during the analysis. The higher proton number of Sn (Z=50) yields an increased cross section for multi-step Coulomb-excitation.

The option of inverse kinematics simplified the targetry and dictated placement of the accompanying detectors. A Micron S3 Dual-sided Silicon Strip Detector (DSSD) was placed downstream of the target ladder to detect the scattered beam and target ions. The detector area spanned $10^{\circ}$ to $28^{\circ}$ in the laboratory frame. The target nuclei were clearly distinguishable from the beam nuclei on the basis of their observed energy and angle, allowing for particle identification. This is demonstrated in Figure \ref{fig:pid} which plots the charge collected as a function of DSSD ring number for $^{200}$Hg scattering on $^{120}$Sn or $^{48}$Ti. Scattering angle ($\theta$) increases with ascending ring number. The active area of the DSSD was subdivided into four angular bins as a compromise between maximising sensitivity to the reorientation effect whilst maintaining an acceptable amount of statistics in each bin. Inner DSSD rings 1 to 8 were excluded from the analysis as the Hg and target loci obscured each other.

The Gammasphere array of high purity germanium (HPGe) \cite{ref:Lee_90} detectors with Compton suppressors surrounded the vacuum chamber containing the targets and DSSD. For this measurement 74 Gammasphere modules were installed. Scattered beam and target nuclei that have undergone Coulomb-excitation de-excited in flight. Following detection by Gammasphere, de-excitation $\gamma$ rays were Doppler corrected to recover their energy in the centre-of-mass frame. The Doppler correction relies on two-body kinematics, the beam energy measurement, and the angle between the ion and time coincident $\gamma$-ray hits. The velocity vector reconstruction uses the segmentation of the DSSD and the known location of the HPGe detectors. The Gammasphere data acquisition system \cite{ref:Anderson_12} utilised a 100 MHz clock speed and ran in a triggered mode requiring a coincident front and back hit in the DSSD. 

The recorded waveforms and time stamps were processed with a modified version of the GEBSort code \cite{ref:Lauritsen}, which itself utilises the \footnotesize{ROOT}\normalsize{} \cite{ref:Brun_97} data format. Figure \ref{fig:gammaspec} shows energy spectra $\gamma$ rays collected in coincidence with scattered target nuclei, correcting for beam and target velocity vectors in black and red respectively. Peaks are labeled with the transition and originating nucleus.

\begin{table}[t]
    \centering
    \begin{footnotesize}
    \begin{tabularx}{\columnwidth}{lllr}
         \hline \hline\\
         $^{200}$Hg & & & Reference\\
         \hline
         $B(E2; 0^+_1 \rightarrow 2^+_1)\quad\quad$ & \small{(e$^2$fm$^4$)}$\quad\quad$ & \small{8530(70)}$\quad$ & \cite{ref:Spear_80} \\

         $B(E2; 2^+_1 \rightarrow 4^+_1)$ & \small{(e$^2$fm$^4$)} & \small{4770(210)} & \cite{ref:Bockisch_79,ref:Gunther_81} \\
         
         $B(E2; 2^+_1 \rightarrow 0^+_2)$ & \small{(e$^2$fm$^4$)} & \small{111(55)} & \cite{ref:Hogenbirk_91} \\
         
         $B(E2; 2^+_1 \rightarrow 2^+_2)$ & \small{(e$^2$fm$^4$)} & \small{150(30)} & \cite{ref:Bockisch_79} \\
         
         $B(M1; 2^+_1 \rightarrow 2^+_2)$ & \small{$(\mu_N^2)$}  & \small{0.0027(7)} & \cite{ref:Subber_87,ref:Breitig_74} \\
         
         $B(E2; 4^+_1 \rightarrow 6^+_1)$ & \small{(e$^2$fm$^4$)} & \small{4600(400)} & \cite{ref:Gunther_81} \\
         
         \hline
         \\
         $^{202}$Hg\\
         \hline \\
         $B(E2; 0^+_1 \rightarrow 2^+_1)$ & \small{(e$^2$fm$^4$)} & \small{6080(50)} & \cite{ref:Spear_80,ref:Bockisch_79} \\

         $B(E2; 2^+_1 \rightarrow 2^+_2)$ & \small{(e$^2$fm$^4$)} & \small{530(180)} & \cite{ref:Bockisch_79} \\
         
         $B(M1; 2^+_1 \rightarrow 2^+_2)$ & \small{$(\mu_N^2)$} & \small{0.0097(23)} & \cite{ref:Kern_19} \\
         
         $B(E2; 2^+_1 \rightarrow 4^+_1)$ & \small{(e$^2$fm$^4$)} & \small{3400(100)} & \cite{ref:Bockisch_79} \\
         
         \hline
         \\
         $^{48}$Ti\\
         \hline \\
         $B(E2; 0^+_1 \rightarrow 2^+_1)$ & \small{(e$^2$fm$^4$)} & \small{720(40)} & \cite{ref:Ernst_00,ref:Bardin_73,ref:Fisher_73} \\

         $Q_s(2^+_1)$ & \small{(eb)} & \small{-0.177(8)} & \cite{ref:Lightbody_72}\\
         
         \hline
         \\
         $^{120}$Sn\\
         \hline \\
         $B(E2; 0^+_1 \rightarrow 2^+_1)$ & \small{(e$^2$fm$^4$)} & \small{1970(30)} & \cite{ref:Hass_80} \\

         $Q_s(2^+_1)$ & \small{(eb)} & \small{-0.05(10)}  & \cite{ref:Leonard_72}\\

        \hline \hline
    \end{tabularx}
    \caption{Literature reduced transition probabilities and spectroscopic quadrupole moments used as constraints. Data were initially sourced from ENSDF evaluations \cite{ref:Kondev_23,ref:Kondev_24,ref:Chen_21,ref:Kitao_02}.}
    \label{tab:lit}
    \end{footnotesize}
\end{table}

For $^{200}$Hg, the $2^+_1 \rightarrow 0^+_1$, $4^+_1 \rightarrow 2^+_1$, $0^+_2 \rightarrow 2^+_1$, $2^+_2 \rightarrow 2^+_1$, and $6^+_1 \rightarrow 4^+_1$ transitions were observed, and the $2^+_1 \rightarrow 0^+_1$, $2^+_2 \rightarrow 2^+_1$, and $4^+_1 \rightarrow 2^+_1$ transitions were observed in $^{202}$Hg. Yields for the $2^+_1 \rightarrow 0^+_1$ transitions were also extracted for $^{48}$Ti and $^{120}$Sn with both Hg isotopes, for the purposes of normalisation. The \footnotesize{JRoot} \normalsize{}\cite{ref:Smallcombe_20} software package was used to fit the experimental peaks. Gamma-ray yields were compared with calculations from the semiclassical coupled channel Coulomb-excitation analysis code \footnotesize{GOSIA} \normalsize{}\cite{ref:Czosnyka_83} executed within an external \footnotesize{C++} \normalsize{}wrapper and minimiser, \footnotesize{GOSIAFitter} \normalsize{}\cite{ref:Henderson_23}. The underlying Coulomb-excitation calculations used in the minimisation procedure included the geometries of both detector systems, level schemes, and the literature values of electromagnetic matrix elements. These were supplemented with internal conversion coefficients calculated with \footnotesize{BRICC} \normalsize{}\cite{ref:Kibedi_08} and stopping powers calculated with \footnotesize{SRIM 2008} \normalsize{}\cite{ref:Ziegler_10} to account for energy loss in the targets.

The minimiser in  \footnotesize{GOSIAFitter} \normalsize utilises the \footnotesize{ROOT MINUIT} \normalsize{} \cite{ref:James_75} libraries. The $\chi^2_{total}$ between the experimental $\gamma$-ray yields and corresponding simulated yields from \footnotesize{GOSIA} \normalsize are minimised through repeated modification to values of the beam and target nuclei matrix elements. Initial values of the matrix elements were derived from the literature $B(E2)$, $B(M1)$, and $Q_s(2^+_1)$ values. Almost all matrix elements that were modified had a literature constraint applied which contributes to $\chi^2_{total}$ (the $\bra{2^+_1}E2\ket{2^+_1}_{Hg}$ matrix element was not literature constrained). The literature values used in this way are shown in Table \ref{tab:lit}. Simultaneous minimisation of the beam and target al.lows for the normalisation of beam and target nuclei, permitting reduced transition probability values to be inferred.

To explore consistency with the previous literature values, the minimisation was performed with and without Hg literature constraints. Results marked as ``unconstrained'' in Table \ref{tab:results} were generated by firstly executing the Hg and Ti \footnotesize{GOSIAFitter} \normalsize simulation with the Hg literature constraints excluded and the Ti literature constraints included. The fitted values and uncertainties of the Hg matrix elements output from the first minimisation procedure were input to the Hg and Sn minimisation as constraints. The $^{120}$Sn literature $B(E2)$ also was a constraint in the second minimisation. The $^{120}$Sn $Q_s(2^+_1)$ when used as a constraint did not have an appreciable impact on the result and was itself well constrained by the present data to a value inconsistent with the value found by Leonard \cite{ref:Leonard_72}. After this was found, the $^{120}$Sn $Q_s(2^+_1)$ was removed as a constraint from all minimisation procedures. The ``constrained'' runs follow the same methodology, however Hg literature constraints were included during the initial Hg and Ti minimisation. The data from another Coulomb-excitation experiment investigating $^{120}$Sn is currently undergoing analysis. The $^{120}$Sn $Q_s(2^+_1)$ value from this work will be compared to the value from this second study once available.

Excitation cross sections are perturbed by the interference terms that arise from the competition between alternative excitation pathways \cite{ref:Zielinska_22}. The case of maximal constructive interference is modelled by a complete set of positive matrix elements. Results of this kind have historically been presented assuming different signs of the $P_3$ quadrupole interference term \cite{ref:Bockisch_79}. Nearby $^{192}$Pt and $^{194}$Pt are two known nuclei with an anomalous positive $P_4$ term, which is linked to the $P_3$ term via the diagonal matrix element $\bra{2^+_1}E2\ket{2^+_1}$ \cite{ref:Allmond_09}. Relative signs of matrix elements for nuclei in this mass region should be carefully considered. The triaxial rotor model can explain $P_3$ or $P_4$ sign anomalies without relying on higher order (hexadecapole) forms of shape deformation \cite{ref:Allmond_09}. Alternative sign configurations of Hg, Sn and Ti matrix elements were simulated. These different simulations resulted in different $\chi^2_{total}$ minima and different final values of the Hg matrix elements. Any run where $\chi^2_{total} \leq \chi^2_{total, min} +1$ was deemed as significant, however the significant sets of Hg matrix elements were quite similar. The uncertainties of the optimal set of Hg matrix elements were increased to account for this which was propagated to the results presented in Table \ref{tab:results} in quadrature.

\begin{table*}[t]
\caption{Results from the present work without (Unconstrained) and with (Constrained) the inclusion of the literature data for the isotopes of interest.}
\label{tab:results}
\begin{footnotesize}
\centering
    \begin{tabularx}{\textwidth}{lXXXXXXr}
        \hline 
        \\ [-15pt]
        \textsuperscript{200}Hg &
        \multicolumn{2}{c}{Unconstrained} &
        \multicolumn{2}{c}{Constrained} &
        \multicolumn{3}{c}{Literature}
        \\
        \hline \\[-15pt]
        $J^\pi_i\rightarrow J^\pi_f$ 
        & $\bra{J^\pi _i}E2\ket{J^\pi_f}$ 
        & $B(E2)\uparrow$
        & $\bra{J^\pi _i}E2\ket{J^\pi_f}$ 
        & $B(E2)\uparrow$
        & $\bra{J^\pi _i}E2\ket{J^\pi_f}$
        & $B(E2)\uparrow$
        & Ref.  \\

        & (eb) & (e$^2$fm$^4$) & (eb) & (e$^2$fm$^4$) & (eb) & (e$^2$fm$^4$) & \\

        \hline \\[-15pt]
        $0^+_1\rightarrow2^+_1$ &
        0.89(6) &
        7970(1000) &
        0.923(4) &
        8530(70) &
        0.924(8) &
        8530(150) &
        \cite{ref:Bockisch_79} \\
        &
        &
        &
        &
        &
        0.924(4) &
        8530(70) &
        \cite{ref:Spear_80} \\
        &
        &
        &
        &
        &
        0.950(38) &
        9540(700) &
        \cite{ref:Olaizola_19} \\
        $2^+_1\rightarrow4^+_1$ &
        1.52(9) &
        4600(600) &
        1.54(3) &
        4800(200) &
        1.527(77) &
        4660(470) &
        \cite{ref:Bockisch_79} \\
        &
        &
        &
        &
        &
        1.55(4) &
        4790(230) &
        \cite{ref:Gunther_81} \\
        $2^+_1\rightarrow0^+_2$ &
        0.24(2) &
        120(20) &
        0.25(2) &
        120(16) &
        0.236(59) &
        111(56) &
        \cite{ref:Olaizola_19}\\
        \hline \\[-15pt]

        $J^\pi_i$ 
        & $\bra{J^\pi _i}E2\ket{J^\pi_i}$
        & $Q_s(J^\pi _i)$
        & $\bra{J^\pi _i}E2\ket{J^\pi_i}$
        & $Q_s(J^\pi _i)$
        & $\bra{J^\pi _i}E2\ket{J^\pi_i}$
        & $Q_s(J^\pi _i)$
        & Ref.  \\

        & (eb) & (eb) & (eb) & (eb) & (eb) & (eb) & \\
        
        \hline \\[-15pt] 
        $2^+_1$ &
        0.5(5) &
        0.4(4) &
        0.7(2) &
        0.5(1) &
        1.47(15) &
        1.11(11) &
        \cite{ref:Bockisch_79} \\
        &
        &
        &
        &
        &
        1.41(25) &
        1.07(19) &
        \cite{ref:Spear_80} \\
        \hline \\[-15pt]
        
        \textsuperscript{202}Hg &
        \multicolumn{2}{c}{Unconstrained} &
        \multicolumn{2}{c}{Constrained} &
        \multicolumn{3}{c}{Literature}
        \\

        \hline \\[-15pt]
        $J^\pi_i\rightarrow J^\pi_f$ 
        & $\bra{J^\pi _i}E2\ket{J^\pi_f}$ 
        & $B(E2)\uparrow$
        & $\bra{J^\pi _i}E2\ket{J^\pi_f}$ 
        & $B(E2)\uparrow$
        & $\bra{J^\pi _i}E2\ket{J^\pi_f}$
        & $B(E2)\uparrow$
        & Ref.  \\

        & (eb) & (e$^2$fm$^4$) & (eb) & (e$^2$fm$^4$) & (eb) & (e$^2$fm$^4$) & \\
        
        \hline \\[-15pt]
        $0^+_1\rightarrow2^+_1$ &
        0.81(6) &
        6600(1000) &
        0.780(3) &
        6080(50) &
        0.785(6) &
        6160(90) &
        \cite{ref:Bockisch_79} \\
        &
        &
        &
        &
        &
        0.778(3) &
        6050(50) &
        \cite{ref:Spear_80} \\
        $2^+_1\rightarrow2^+_2$ &
        0.5(1) &
        500(200) &
        0.51(1) &
        526(20) &
        0.515(9) &
        530(180)  &
        \cite{ref:Bockisch_79} \\
        &
        &
        &
        &
        &
        0.412(42) &
        340(70) &
        \cite{ref:Agarwal_85}\\
        $2^+_1\rightarrow4^+_1$ &
        1.4(1) &
        3700(600) &
        1.30(2) &
        3400(10) &
        1.311(24) &
        3438(126) &
        \cite{ref:Bockisch_79} \\
        \hline \\[-15pt]
        
        $J^\pi_i$ 
        & $\bra{J^\pi _i}E2\ket{J^\pi_i}$
        & $Q_s(J^\pi _i)$
        & $\bra{J^\pi _i}E2\ket{J^\pi_i}$
        & $Q_s(J^\pi _i)$
        & $\bra{J^\pi _i}E2\ket{J^\pi_i}$
        & $Q_s(J^\pi _i)$
        & Ref.  \\

        & (eb) & (eb) & (eb) & (eb) & (eb) & (eb) & \\
        
        \hline \\[-15pt] 
        $2^+_1$ &
        1.0(6) &
        0.8(4) &
        0.8(1) &
        0.6(1) &
        0.42(18) &
        0.32(14) &
        \cite{ref:Bockisch_79} \\
        &
        &
        &
        &
        &
        1.33(17) &
        1.01(13) &
        \cite{ref:Spear_80} \\
        \hline 
    \end{tabularx}
    \end{footnotesize}
\end{table*}

\section{Discussion}

The results presented in Table \ref{tab:results} offer independent measurements of the $B(E2;0^+_1\rightarrow2^+_1)$ and $Q_s(2^+_1)$ of $^{200}$Hg and $^{202}$Hg. The $B(E2)$ values produced from this work, both Hg literature constrained and unconstrained, are consistent with the literature values. In contrast, the $Q_s(2^+_1)$ values show a significant departure from the literature values.

For $^{200}$Hg, the constrained $Q_s(2^+_1)$ from this work (0.52(14) e$.$b) is $\sim4\sigma$ below the Bockisch \textit{et al.} (1.11(11) e$.$b) \cite{ref:Bockisch_79} and Spear \textit{et al.} (1.07(19) e$.$b) \cite{ref:Spear_80} values. The equivalent unconstrained value has a much larger uncertainty, but still confirms the oblate nature of the deformation and is inconsistent with the literature. As for $^{202}$Hg, the constrained $Q_s(2^+_1)$ (0.63(10) e$.$b) and the values from Bockisch \textit{et al.} (0.32(14) e$.$b) \cite{ref:Bockisch_79} and Spear \textit{et al.} (1.01(13)) \cite{ref:Spear_80} are all inconsistent with one another. The value from this work is $\sim2\sigma$ above the Bockisch \textit{et al.} value and $\sim2\sigma$ below the Spear \textit{et al.} value \cite{ref:Spear_80}. The unconstrained

\begin{figure}[h!]
    \centering
    \includegraphics[width=1.\linewidth, trim=0.3cm 0.5cm 0.4cm 0.7cm, clip]{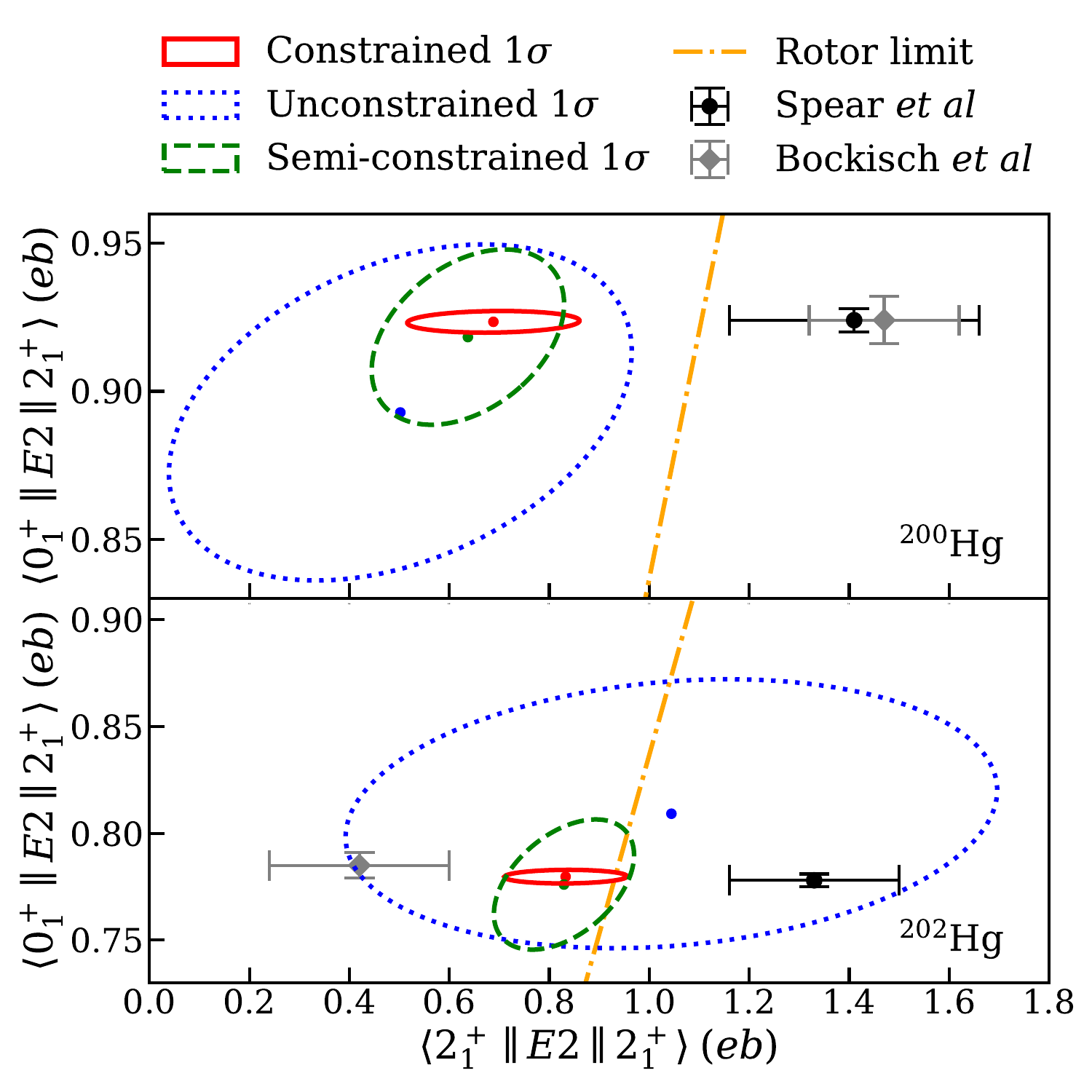}
    \caption{Confidence intervals determined via covariance matricies from \footnotesize{GOSIAFitter} plotted with previous literature values and the rigid rotor value. Results from this work are shown fully literature constrained, semi-constrained and unconstrained. Here, semi-constrained means that all Hg literature data were included except $B(E2)$ values. Data to the right of the dash-dotted rotor line exceed the limit.}
    \label{fig:covar}
\end{figure}

\begin{figure*}[th!]
    \centering
    \includegraphics[width=0.95\linewidth, trim=0.2cm 0.2cm 0.2cm 0.2cm, clip]{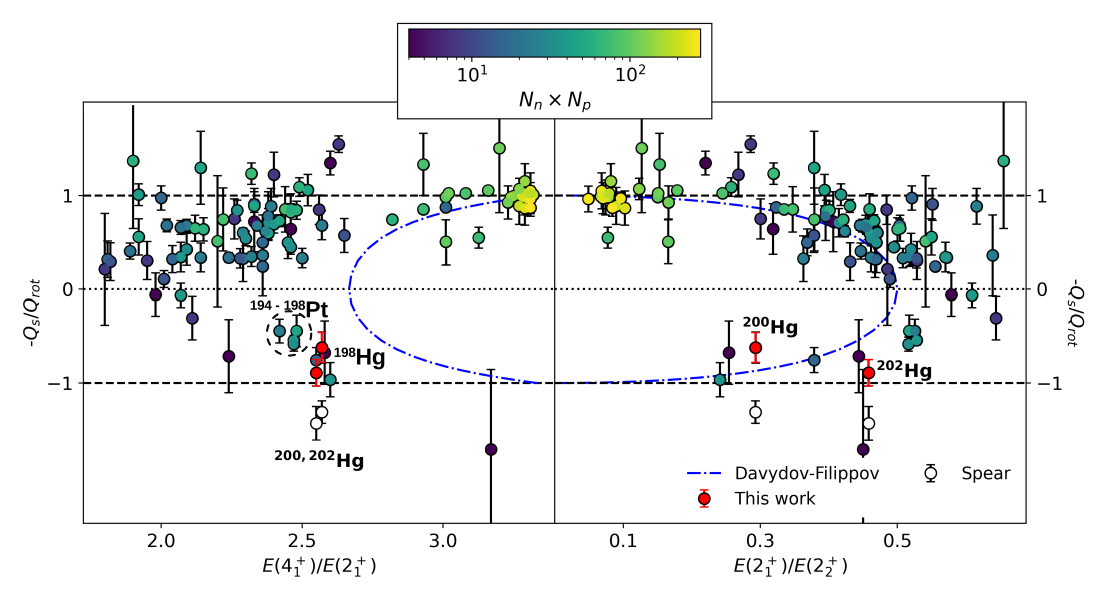}
    \caption{The ratio of the spectroscopic quadrupole moment to rotor model prediction is shown against energy ratios $E_{4^+_1}/E_{2^+_1}$, and $E_{2^+_1}/E_{2^+_2}$. The colour-scale indicates the product of the number of valence neutrons (or neutron holes) and number of valence protons (or proton holes) in the nucleus ($N_n\times N_p$), with the most deformed, axial systems typically corresponding to those with the largest $N_n\times N_p$ value. Mercury and platinum isotopes are highlighted, indicating their unusual status as well-deformed, oblate systems. Present results for \textsuperscript{200,202}Hg bring them into line with the physically allowed space of $\left|Q_s/Q_{rot}\right|\leq1$. }
    \label{fig:systematics}
\end{figure*}

\noindent $Q_s(2^+_1)$ is also not consistent with the literature values. The 1$\sigma$ confidence intervals from covariances of the matrix elements produced from \footnotesize{GOSIAFitter} \normalsize{} minimisations are plotted in Figure \ref{fig:covar}. Alongside this are the Bockisch \textit{et al.} \cite{ref:Bockisch_79} and Spear \textit{et al.} \cite{ref:Spear_80} $Q_s(2^+_1)$ values and the limit for a Bohr-Mottelson rigid rotor. A third set of \footnotesize{GOSIAFitter} \normalsize{} results (``semi-constrained'') are also presented which utilised all available Hg literature as constraints except for the $B(E2; 0^+_1\rightarrow 2^+_1)$. The semi-constrained results demonstrate the $Q_s(2^+_1)$ values from this work are correlated but are not dependent on the $B(E2)$ value, in case the $B(E2)$ is remeasured in the future.

Without exception, the unconstrained values with their uncertainty are consistent with the equivalent constrained values. All unconstrained $B(E2)$ values are consistent with the corresponding literature values, although the uncertainties are sufficiently large that this is not challenging. In contrast to this, the new constrained and unconstrained $Q_s(2^+_1)$ values for both $^{200}$Hg and $^{202}$Hg show a significant departure from the values of Bockisch \textit{et al.}: 1.11(11), 0.32(14) e$.$b \cite{ref:Bockisch_79}, and Spear \textit{et al.}: 1.07(19), 1.01(13) e$.$b \cite{ref:Spear_80}.

The most stark methodology change from the Bockisch \textit{et al.} and Spear \textit{et al.} experiments was the shift to inverse kinematics and Hg beams. The use of Ti and Sn as targets provided a more collective reaction partner with a higher proton number than the $^{4}$He, $^{12}$C, $^{16}$O light ion beams previously used, hence the cross section of multi-step Coulomb-excitation was higher and the reorientation effect was more significant. 

This, combined with the use of Gammasphere for greater $\gamma$-ray detection efficiency, plus the ample spectroscopic information available on $^{48}$Ti and $^{120}$Sn produces a better constrained experimental system. The use of a segmented silicon detector also brought sensitivity to the angular distribution of the target recoils, although the target normalisation as a means of accessing the absolute cross section was responsible for the majority of the sensitivity to $Q_s(2^+_1)$. Coulomb-excitation cross sections are only perturbed by $Q_s(2^+_1)$ \cite{ref:Zielinska_22}. The reduced transition probabilities carry the majority of the influence, hence the improvements made to the setup were necessary to improve the sensitivity to $Q_s(2^+_1)$. The significance of the new Hg $Q_s(2^+_1)$ values are demonstrated in Figure \ref{fig:systematics} where both energy ratios $E(4^+_1)/E(2^+_1)$ and $E(2^+_1)/E(2^+_2)$ are plotted against $-Q_s(2^+_1)/Q_{rot}(2^+_1)$. The value $Q_{rot}(2^+_1)$ is the idealised axial rotor model value, calculated from the $B(E2)$. The above ratio of experimental and theoretical values of $Q$ is the reduced quadrupole moment. Nuclei are expected to lie outside of the dashed-dotted blue Davydov-Filippov envelope.

Utilising the Spear \textit{et al.} \cite{ref:Spear_80} $Q_s(2^+_1)$ values, both $^{200}$Hg and $^{202}$Hg fell outside the $Q_{reduced}=-1$ limit. The new data points (red) calculated from the constrained $Q_s(2^+_1)$ values are more in line with neighbouring nuclei, $^{198}$Hg and $^{194,196,198}$Pt. It must be noted that the $^{200}$Hg datum when plotted against $E(2^+_1)/E(2^+_2)$ falls outside of the boundary produced with the Davydov-Filippov triaxial rotor model \cite{ref:Davydov_59}. This unidealised rotor behaviour could be evidence of configuration mixing in $^{200}$Hg between the $2^+$ states. It is also suspected that shape coexistence occurs in both $^{200,202}$Hg due to the presence of low lying $0^+_2$ states at 1029 and 1411 keV, respectively.

\begin{figure}[h!]
    \centering
    \includegraphics[width=1.\linewidth, trim=0.4cm 0.5cm 0.3cm 0.1cm, clip]{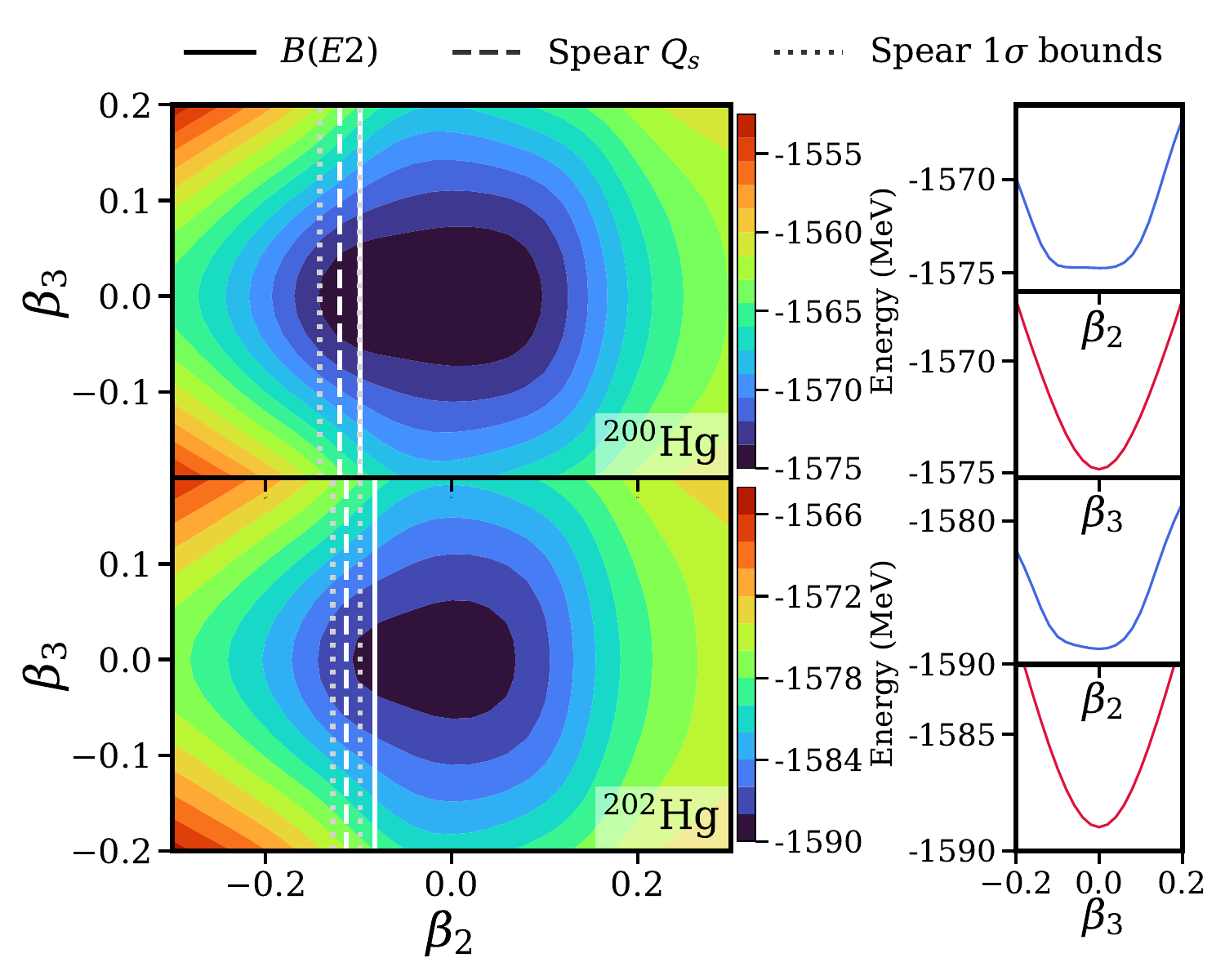}
    \caption{Potential energy surface calculation results using SkX force. The $\beta_2$ values calculated from the Spear \textit{et al.} $Q_s(2^+_1)$ \cite{ref:Spear_80} and the evaluated $B(E2)$ values \cite{ref:Kondev_23,ref:Kondev_24} have been added. The uncertainty in the $B(E2)$ derived $\beta_2$ value is too small to visualise. The comparison excludes prolate (positive $\beta_2$) deformation because $^{200,202}$Hg are known to be oblate by their measured $Q_s(2^+_1)$ moments. Projections are also shown of $\beta_2$ along $\beta_3=0$ and of $\beta_3$ along $\beta_2=0$.}
    \label{fig:pes}
\end{figure}

Potential energy surface calculations were also carried out for $^{200}$Hg and $^{202}$Hg to analyse the shape of these nuclei further \cite{ref:Reinhard_21}. The results of these are shown in Figure \ref{fig:pes} as a function of the $\beta_2$ and $\beta_3$ deformation parameters. For both $^{200}$Hg and $^{202}$Hg, the global minima are at $\beta_2=\beta_3=0$, but with softness towards the negative $\beta_2$ direction. This suggests a multiconfiguration wave function with an averaged $\beta_2$ value below zero. The calculations predict a greater degree of softness in both deformation parameters for $^{200}$Hg as compared to $^{202}$Hg, as shown by the wider energy minimum for $^{200}$Hg in Figure \ref{fig:pes}. The lack of observation of $\gamma$ rays from $3^-$ states in this work is consistent with the small $\beta_3$ from the calculations.

\section{Conclusions}

A Coulomb-excitation measurement utilising inverse kinematics and a high-Z beam was performed to remeasure the spectroscopic quadrupole moment $Q_s(2^+_1)$ of $^{200}$Hg and $^{202}$Hg. The previous literature $Q_s(2^+_1)$ values from Bockisch \textit{et al.} \cite{ref:Bockisch_79} and Spear \textit{et al.} \cite{ref:Spear_80} were unexpectedly large for $^{200}$Hg, and were inconsistent with one another for $^{202}$Hg by $\sim5\sigma$. Results marked as unconstrained were obtained independently of the Hg literature. The $^{200}$Hg $Q_s(2^+_1)$ was found to be 0.38(35) or 0.52(14) eb when unconstrained or constrained respectively. The $^{202}$Hg $Q_s(2^+_1)$ was found to be 0.79(44) or 0.63(10) eb when unconstrained or constrained respectively. The $Q_s(2^+_1)$ values from this work are more typical considering the systematics of neighbouring nuclei in the $A\approx200$ region. The values now also have a smaller magnitude than the theoretical limit of a quadrupole deformed rigid rotor, which shows these nuclei are not total axial rotors. Furthermore, triaxiality or configuration mixing may be occuring. The consequence of smaller $Q_s(2^+_1)$ values in conjunction with the $\beta_3 \approx0$ results from the potential energy surface calculations is a reduction in the magnitude of the Schiff moment for nuclei in this region. Due to the quadratic dependency of the Schiff moment on $\beta_3$, $B(E3; 0^+_1\rightarrow3^-_1)$ measurements via state lifetimes should be performed to confirm understanding of the exotic deformation in this region.

\section*{Acknowledgements}
The authors thank the ATLAS operational staff of Argonne National Laboratory for providing the beam. Work at the University of Surrey was supported under UKRI Future Leaders Fellowship (grants no. MR/T022264/1 and MR/Y034007/1) and the Science and Technologies Facilities Council (STFC). Work of Argonne National Laboratory was supported by the U.S. Department of Energy, Office of Science, Office of Nuclear Physics, under contract number DE-AC02-06CH11357. Work at Lawrence Livermore National Laboratory was performed under the auspices of the U.S. Department of Energy under contract DE-AC52-07NA27344. This research uses resources of ANL’s ATLAS facility, which is a U.S. Department of Energy Office of Science User facility.